\documentclass[10pt,journal,compsoc]{IEEETran}

\usepackage{bbold}
\usepackage{amsmath}
\usepackage{graphicx}

\newtheorem{definition}{Definition}

\usepackage{multirow}
\usepackage[pagebackref=true,breaklinks=true,letterpaper=true,colorlinks,bookmarks=false]{hyperref}

\usepackage[dvipsnames]{xcolor}
\usepackage[normalem]{ulem}
\definecolor{mred}{rgb}{.80,.12,.30}
\definecolor{MRED}{rgb}{.80,.12,.30}
\definecolor{grey}{rgb}{0.5,0.5,0.5}
\definecolor{lgrey}{rgb}{0.7,0.7,0.7}
\definecolor{purple}{rgb}{.75,0,.85}
\definecolor{cyan}{rgb}{0,0.68,.94}
\definecolor{pistachio}{rgb}{0.58, 0.77, 0.45}
\definecolor{myorange}{rgb}{0.94, 0.36, 0.13}

\newif\ifnotes
\notestrue

\begin{document}

\title{Modeling Spatial Nonstationarity via Deformable Convolutions for Deep Traffic Flow Prediction}

\author{Wei~Zeng,~\IEEEmembership{Member,~IEEE},~Chengqiao Lin,~Kang Liu,\\~Juncong~Lin,~\IEEEmembership{Member,~IEEE},~Anthony K. H. Tung,~\IEEEmembership{Senior Member,~IEEE}
\IEEEcompsocitemizethanks{
\IEEEcompsocthanksitem
W. Zeng and K. Liu are with Shenzhen Institute of Advanced Technology, Chinese Academy of Sciences. e-mail: \{wei.zeng, kang.liu\}@siat.ac.cn.
\IEEEcompsocthanksitem
C. Lin and J. Lin (corresponding author) are with Xiamen University, China. E-mail: \{linchengqiao, jclin\}@xmu.edu.cn.
\IEEEcompsocthanksitem
A.K.H. Tung is with School of Computing, National University of Singapore. E-mail: atung@comp.nus.edu.sg.
}
\thanks{Manuscript received xx xx, 202x; revised xx xx, 202x.}
}

\markboth{IEEE Transactions on Knowledge and Data Engineering}%
{Zeng \MakeLowercase{\textit{et al.}}: Modeling Spatial Nonstationarity via Deformable Convolutions for Deep Traffic Flow Prediction}

\newcommand{\red}[1]{{\color{red}{#1}}}
\newcommand{\zw}[1]{{\color{red}{#1}}}
\newcommand{\lcq}[1]{{\color{blue}{#1}}}
\newcommand{\blue}[1]{{\color{blue}{#1}}}
\newcommand{\eg}{\emph{e.g.}}
\newcommand{\ie}{\emph{i.e.}}


\IEEEtitleabstractindextext{

\begin{abstract}
Deep neural networks are being increasingly used for short-term traffic flow prediction, which can be generally categorized as convolutional (CNNs) or graph neural networks (GNNs).
CNNs are preferable for region-wise traffic prediction by taking advantage of localized spatial correlations, whilst GNNs achieves better performance for graph-structured traffic data.
When applied to region-wise traffic prediction, CNNs typically partition an underlying territory into grid-like spatial units, and employ standard convolutions to learn spatial dependence among the units.
However, standard convolutions with fixed geometric structures cannot fully model the nonstationary characteristics of local traffic flows.
To overcome the deficiency, we introduce deformable convolution that augments the spatial sampling locations with additional offsets, to enhance the modeling capability of spatial nonstationarity.
On this basis, we design a deep deformable convolutional residual network, namely \emph{DeFlow-Net}, that can effectively model global spatial dependence, local spatial nonstationarity, and temporal periodicity of traffic flows.
Furthermore, to better fit with convolutions, we suggest to first aggregate traffic flows according to pre-conceived regions or self-organized regions based on traffic flows, then dispose to sequentially organized raster images for network input.
Extensive experiments on real-world traffic flows demonstrate that \emph{DeFlow-Net} outperforms GNNs and existing CNNs using standard convolutions, and spatial partition by pre-conceived regions or self-organized regions further enhances the performance.
We also demonstrate the advantage of \emph{DeFlow-Net} in maintaining spatial autocorrelation, and reveal the impacts of partition shapes and scales on deep traffic flow prediction.
\end{abstract}

}
\maketitle
\section{Introduction}
\label{sec:intro}

\IEEEPARstart{T}{raffic} flow prediction should yield accurate projections on the expected traffic conditions, in order to support intelligent transportation systems~\cite{vlahoginani_2004_short-term}.
Numerous data-driven approaches, such as auto regressive integrated moving average (ARIMA) and its variants (\eg,~\cite{moorthy_1988_short, willianms_1998_urban}) that take advantages of repeating occurrences in temporal historical traffic data, have been developed for traffic flow prediction.
However, it is a challenging task for conventional approaches to model the complex non-linear spatial and temporal patterns of traffic flows.
Recently, research focus has shifted towards utilizing deep neural networks (DNNs) for traffic flow prediction.
Many DNN-based solutions have been developed, including residual convolution neural networks (CNNs) that are preferable for region-wise traffic flow prediction, and graph neural networks (GNNs) for graph-structured traffic data, \eg, flow volumes on road and subway networks. Studies have revealed superior performances of DNNs than conventional approaches.

Specifically, CNN approaches typically partition an underlying territory into grid-like regions, and aggregate in- and out-flows in each region.
In this way, traffic flows are represented in raster image format, which is consumable by convolutions.
Next, standard convolutions are employed to learn the spatial dependence of traffic flows between pairs of locations.
Standard convolution has shown to be effective, since traffic flows observed at a spatial unit are dependent on traffic flows at nearby spatial units, which is called \emph{spatial dependence}~\cite{tobler1970computer}.
In a context of urban environments, human mobility flows are highly associated with the functionality of a place~\cite{yuan_2012_discovering, zeng_2017_visualizing, feng_2021_topology}.
For example, two residences may show similar periodic patterns, \eg, large out-flows in the morning and in-flows in the evening on weekdays.
In contrast, daily pendulum movements can be observed between residential and official places~\cite{zeng_2015_visualizing}.

\begin{figure}
  \centering
  \includegraphics[width=0.495\textwidth]{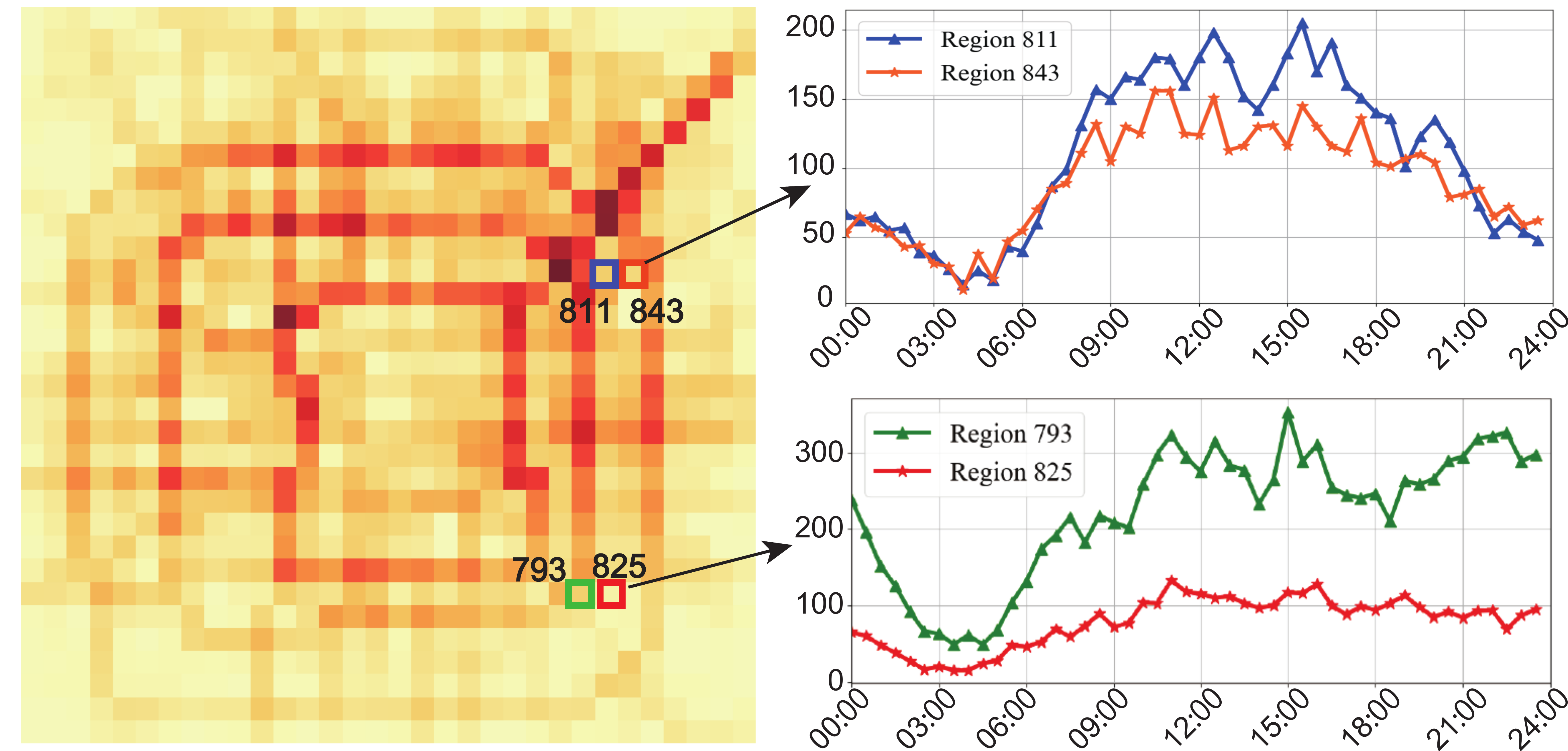}
  \vspace{-5mm}
  \caption{Traffic flows intrinsically feature locally spatial nonstationarity between sets of neighboring regions. Traffic flows of regions 811 \& 843 show similar temporal variations (right-top), whilst those of regions 793 \& 825 are distinct (right-bottom).}
  \vspace{-3mm}
  \label{fig:Spatial_dependence}
\end{figure}

Besides spatial dependence, traffic flows also feature locally \emph{spatial nonstationarity} that depicts varying relationships between some sets of variables over space~\cite{brunsdon1996geographically}.
Fig.~\ref{fig:Spatial_dependence} (left) presents traffic flows in Beijing under a 32$\times$32 grid divison.
Two sets of neighboring regions are selected: 1) regions 811 \& 843, and 2) regions 793 \& 825.
Temporal variations of averaged traffic flows of these regions are presented in Fig.~\ref{fig:Spatial_dependence} (right).
Here, regions 811 \& 843 show similar flow dynamics, whilst those of regions 793 \& 825 are different.
Standard convolution with a fixed geometric structure (\eg, a 3$\times$3 kernel) cannot capture the varying relationships between the two sets of regions, not to mention if we consider sets of over two regions.
As such, existing convolution-based approaches are incompetent to model spatial nonstationarity of traffic flows.

This paper tackles the problems with a feasible solution from two perspectives.
First, we observe that most of existing convolution-based approaches for traffic flow prediction partition the underlying territory into grid regions.
The partition scheme however increases spatial nonstationarity, and consequently harms prediction performance of the network.
To mitigate the issue, we suggest to aggregate traffic flows according to pre-conceived regions (\eg, traffic analysis zones (TAZs) or functional regions), or self-organized regions based on traffic flows.
The traffic flows aggregated in irregular regions are passed through a rasterization process and disposed to raster images.
Second, we propose to advocate the use of deformable convolution~\cite{dai2017deformable}, which has shown to be effective for many applications, such as semantic image segmentation~\cite{CHEN2021853} and image deblurring~\cite{yuan_2020_efficient}.
Specifically, deformable convolution augments the spatial sampling locations with additional offsets, and learns the offsets from the spatial distributions of traffic flows.
In this way, deformable convolution not only inherits the ability of standard convolution in learning spatial dependence, but also being able to alter over space to capture the nonstationary characteristics of traffic flows.

On these basis, we design \emph{DeFlow-Net}, which is a deformable convolutional residual network for accurate short-term traffic flow prediction.
Besides deformable convolutions for modeling spatial dependence and nonstationarity, \emph{DeFlow-Net} also incorporates a temporal dependency module to learn temporal patterns of traffic flows from weekly trend, daily periodicity, and hourly closeness components.
The temporal feature components are fused together to predict the future trend.
We conduct extensive experiments on real-world traffic flows in Beijing, New York City, and Shenzhen.
The results demonstrate that \emph{DeFlow-Net} outperforms existing CNN and GNN methods for region-wise traffic flow prediction, and ablated techniques with standard and atrous convolutions, in terms of both prediction accuracy and the ability to maintain spatial autocorrelation.
The study further reveals that spatial partition by pre-conceived or self-organized regions also contributes to performance improvements.

The main contributions of this work include:
\begin{itemize}
\item
We design \emph{DeFlow-Net}, a deep deformable convolutional residual network that takes advantages of deformable convolutions to model both spatial dependence and nonstationarity (Sect.~\ref{ssec:deflow-net}).
To the best of our knowledge, it is the first effort that attempts to model both spatial dependence and nonstatinarity in a deep-learning-based model for traffic prediction.

\item
We evaluate the effectiveness of \emph{DeFlow-Net} on real-world traffic flows in three cities, and compare with baseline models and ablated techniques.
The comparison results indicate our proposed model outperforms existing methods in terms of both prediction accuracy (Sect.~\ref{ssec:Result}) and the ability to preserve spatial autocorrelation (Sect.~\ref{ssec:spatial_analysis}).

\item
We conduct local indicators of spatial autocorrelation (LISA) analysis~\cite{anselin1995local}, and reveal the impacts of partition shapes and scales on predictions (Sect.~\ref{ssec:diffshape}).
The study confirms the effectiveness of integrating pre-conceived or self-organized regions (Sect.~\ref{ssec:input}), and also the effectiveness of conventional spatial autocorrelation analysis (Sect.~\ref{ssec:spatial_analysis}), when preparing network inputs.

\end{itemize}
\begin{table}[!htb]
    \centering
   \vspace{-2mm}
    \caption{Meanings of all notations.} 
    \vspace{-2mm}
    \resizebox{0.995\linewidth}{!}{
        \begin{tabular}{|c|l|} \hline
        \textbf{SYMBOL} & \textbf{DESCRIPTION}  \\ \hline\hline
        $\mathcal{M}; m$ & all movements; a movement. \\ \hline
        $\mathcal{T}; t$ & all time slots; a time slot. \\ \hline

        $\mathcal{R}; r$ & all regions; a region. \\ \hline 
        $X_{\mathcal{R},t}$; & in-/out-flow of regions $\mathcal{R}$ at time slot $t$; \\
        $x_{r,t}$ & in-/out-flow of a region $r$ at time slot $t$.\\ \hline \hline

        $G; g$ & grid map; a grid.\\ \hline
        $X_{G,t}$; & in-/out-flow of grid map $G$ at time slot $t$; \\
        $x_{g,t}$ & in-/out-flow of a grid $g$ at time slot $t$. \\ \hline
        $Y_{G,t}$; & predicted in-/out-flow of grid map $G$ at time slot $t$; \\
        $y_{g,t}$ & predicted in-/out-flow of a grid $g$ at time slot $t$. \\ \hline
        \end{tabular}
    }
  \vspace{-3mm}
    \label{table:notations}
\end{table}

\section{Problem Definition}
\label{sec:problem}

This section introduces fundamental concepts of our work. 
Table~\ref{table:notations} lists down the notations used throughout the paper.
The input data in our problem is a set of movements $\mathcal{M}$ that denotes the mobility of multiple moving objects (\eg, mobile users and taxi cabs) throughout a time period $\mathcal{T}$.
We divide $\mathcal{T}$ into equal-time slots to analyze the movements in time-series discretely, \emph{i.e.}, $\mathcal{T} := \left\{t_1, \cdots, t_n\right\}$.

\begin{definition}[\textbf{Movement}]
\label{def:move}
A \emph{movement} $m$ is a continuously measured trajectory of a moving object during a time period $\mathcal{T}$, which is defined by a set of spatiotemporal records $\bigcup_{t \in \mathcal{T}} <t, l_t>$, where $l_t$ represents the position of $m$ at time $t$. 
We denote all movements of multiple moving objects as $\mathcal{M}$.
\end{definition}

\begin{definition}[\textbf{Region}]
\label{def:region}
We partition an underlying territory into a set of regions $\mathcal{R}$ that are non-overlapping and fill-up the territory.
Each \emph{region} $r$ can be 1) a grid partitioned based on the longitude and latitude, 2) a pre-conceived region, \eg, a functional zone isolated by road network, or a TAZ constructed by census block information for tabulating traffic-related data, or 3) a self-organized region based on  interaction of traffic flows (\eg,~\cite{2018Capturing, anwar2014spatial, 2016Tracking, anwar2017partitioning}) or similarity of human activities (\eg,~\cite{yuan_2012_discovering, yuan_2015_discovering}). 
Pre-conceived and self-organized regions are typically in irregular shapes.
\end{definition}

From the movements $\mathcal{M}$, we compute inflow $x^{I}_{r, t}$ and outflow $x^{O}_{r, t}$ per time slot $t$ for each region $r \in \mathcal{R}$.

\begin{definition}[\textbf{Inflow \& Outflow}]
We define the inflow $x^I_{r, t}$ of a region $r$ at time slot $t$ as the number of moving objects who are not in $r$ at time slot $t-1$ and appear in $r$ at time slot $t$, \ie,
\begin{equation}
x^I_{r, t} = \vert \left\{m \in \mathcal{M} | m.l_{t-1} \notin r \wedge m.l_t \in r\right\} \vert,
\end{equation}
where $\vert \cdot \vert$ denotes the cardinality of the set.
Similarly, we can compute the outflow $x^O_{r, t}$ as
\begin{equation}
x^O_{r, t} = \vert \left\{m \in \mathcal{M} | m.l_{t-1} \in r \wedge m.l_t \notin r\right\} \vert.
\end{equation}   

\end{definition}

For simplicity, we omit the notations $I/O$ in the following.
The simplified symbols represent either inflow or outflow, unless specified.
We compute in/out traffic flows for all regions $r \in \mathcal{R}$ at all time slots $t \in \mathcal{T}$, yielding a series of traffic flows $\left\{ X_{\mathcal{R}, t} | t \in \mathcal{T} \right\}$.

\vspace{2mm}
\noindent
\textbf{Problem Definition (Short-Term Traffic Flow Prediction)}.
Consider a set of movements $\mathcal{M}$ in the time duration of $\mathcal{T}$, a set of regions $\mathcal{R}$, and a series of derived traffic flows $\left\{ X_{\mathcal{R}, t} | t \in \mathcal{T} \right\}$, our goal is to predict unobserved traffic flows $Y_{\mathcal{R}, t_{n+1}}$ for $\mathcal{R}$ at time slot $t_{n+1}$.
Accuracy is of primary concern for short-term traffic flow prediction, \ie, output predictions shall be close to ground truths.
In addition, this work also considers \emph{spatial autocorrelation} as a key performance indicator, as spatial units of locally high nonstationarity are more likely to produce high prediction errors~\cite{zeng2020revisiting}.

\section{Methodology}
\label{ssec:network}

We propose to tackle the problem from two perspectives.
First, we introduce a preprocessing procedure that integrates pre-conceived or self-organized regions when preparing network consumable inputs (Sect.~\ref{ssec:input}). 
Next, we present \emph{DeFlow-Net}, a deep deformable convolutional residual neural network built upon deformable convolutions (Sect.~\ref{ssec:deflow-net}).
The training scheme for \emph{DeFlow-Net} is described in Sect.~\ref{ssec:training}.

\subsection{Data Processing}
\label{ssec:input}

As discussed in Definition~\ref{def:region}, this work considers both grid partitions based on the longitude and latitude, and partitions by irregular pre-conceived regions such as TAZs or self-organized regions by traffic flows.
The alternative partition approaches can generate different traffic flows, as statistical measurements are subject to the boundaries of spatial units, \ie, \emph{the modifiable areal unit problem (MAUP)}~\cite{gehike_1934_certain, openshaw_1984_modifiable}.
In addition, studies have shown that deep neural networks generally suffer from the adversarial perturbation problem~\cite{Moosavi_2017_universal, zheng_2016_improving}.
As such, the network predictions of traffic flows can be significantly different, even if the input difference is marginal.
However, most of existing convolution-based traffic prediction methods only explore grid partitions based on the longitude and latitude, \eg,~\cite{zhang_2017_deep, yao2018deep}, whilst neglect partitions by pre-conceived or self-organized regions.
This work examines the impacts of different partition methods on convolution-based network predictions. 

\begin{itemize}

\item
\textit{Grid partition}:
First, the longitude and latitude ranges of the studying area are identified.
Next, the studying area is divided into grids of proper size.
For example, TaxiBJ dataset partitions Beijing into 32$\times$32 grids, BikeNYC dataset partitions NYC into 8$\times$16 grids, and TaxiSZ dataset partitions Shenzhen into 50$\times$25, 100$\times$50, and 200$\times$100 grids.  

\item
\textit{Pre-conceived regions}: 
The partition scheme integrates some subjective background knowledge to specify a set of regions.
A typical example is traffic analysis zones (TAZs) constructed by census block information for tabulating traffic-related data.
TAZs are used, for instance, in constructing origin-destination matrices, a classical tool of transportation engineering for describing traffic flows.
Pre-conceived regions can also be incorporated to prepare traffic flows for network prediction, \eg,~\cite{zeng2020revisiting}.

\item
\textit{Self-organized regions}:
There are however certain limitations for pre-conceived regions:
1) the regions are typically formed in a long time ago that may not accord with traffic flows;
and 2) the number of regions are prefixed and lack of flexibility for multi-scale analysis.
Alternatively, one can construct self-organized regions based on traffic flows (\eg,~\cite{2018Capturing, anwar2014spatial, 2016Tracking, anwar2017partitioning}), or human activities (\eg,~\cite{yuan_2012_discovering, yuan_2015_discovering}).

To validate our model upon self-organized regions, we adopt a similarity based community division algorithm measured by traffic-flow interactions between adjacent regions.
Specifically, the traffic-flow interaction $\operatorname{S}\left(r_{i}, r_{j}\right)$ is calculated as:
\begin{equation}
\label{eq:similarity}
\operatorname{S}\left(r_{i}, r_{j}\right)=\frac{2 \times \left(x_{r_{i} \rightarrow r_{j}}+ x_{r_{j} \rightarrow r_{i}}\right)}{x^I_{r_i}+x^O_{r_i}+x^I_{r_j}+r^O_{r_j}}
\end{equation}
where $x_{r_{i} \rightarrow r_{j}}$ denotes the number of all-time flows from region $r_{i}$ to region $r_{j}$, and $x^I_{r_i}$ denotes the number of all-time inflow for region $r_{i}$.
Note that $r_i$ and $r_j$ must meet the condition of being adjacent.
Next, we merge the adjacent regions $r_{i}$ and $r_{j}$ if $\operatorname{S}\left(r_{i}, r_{j}\right)$ exceeds a predefined threshold $\alpha$.
A large $\alpha$ will yield more number of regions than a small $\alpha$.
The process is repeated iteratively until no more regions can be merged.
\end{itemize}

\begin{figure}
  \centering
  \includegraphics[width=0.495\textwidth]{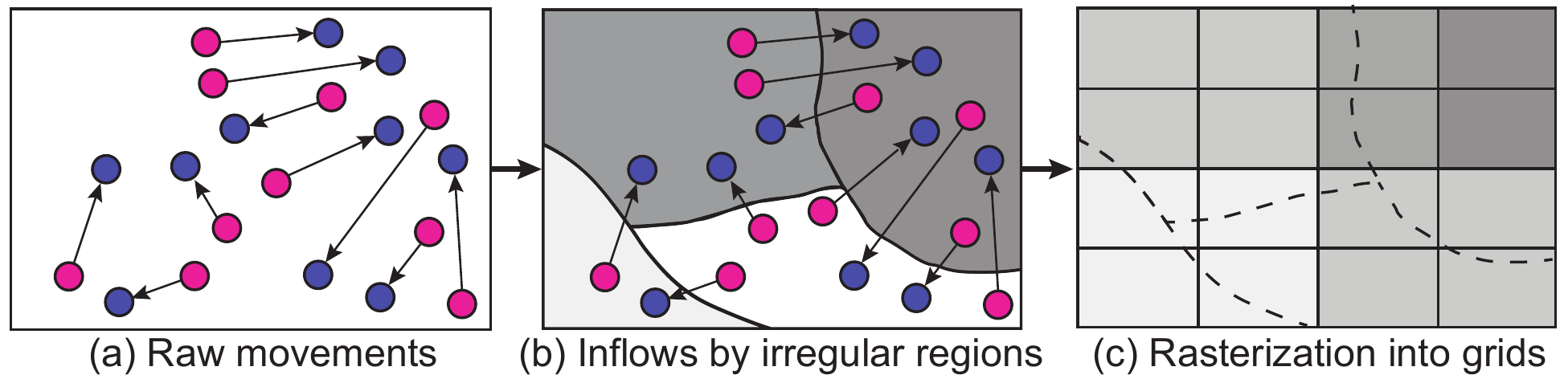}
  \vspace{-7mm}
  \caption{A general pipeline of creating gridded traffic flows as network inputs: Raw movements (a) are counted as inflows based on irregular regions (b), and further rasterized into a grid map (c).}
  \vspace{-4mm}
  \label{fig:raster}
\end{figure}

Using pre-conceived or self-organized regions in irregular shapes, the generated in/out traffic flows $X_{\mathcal{R},t}$ show non-grid-like topology.
However, a convolution operation needs inputs to be in raster-image format which can by processed by the receptive field.
As such, we first convert $X_{\mathcal{R}, t}$ into a raster image $X_{G, t} \in \mathbb{R}^{i\times j}$.
The process is denoted as \emph{rasterization}, as illustrated in Fig.~\ref{fig:raster}.

\begin{definition}[\textbf{Rasterization}]
We divide the underlying territory into a grid map $G$ of size $i \times j$ (\eg, $4 \times 4$ as in Fig.~\ref{fig:raster}).
Each grid $g \in G$ can intersect with arbitrary number of irregular regions $\{r_k\}_{k=1}^n$.
We calculate the in/out traffic flow for each grid $g$ at time slot $t$ as:
\begin{equation}
x_{g,t} = \sum_{k=1}^n x_{r_k,t} \times \frac{S(r_k \cap g)}{S(r_k)},
\label{equ:raster}
\end{equation} 
where $S(\cdot)$ stands for the area of a region, and $r_k \cap g$ indicates the intersection between $r_k$ and $g$.

\end{definition}

\begin{figure}
  \centering
  \includegraphics[width=0.495\textwidth]{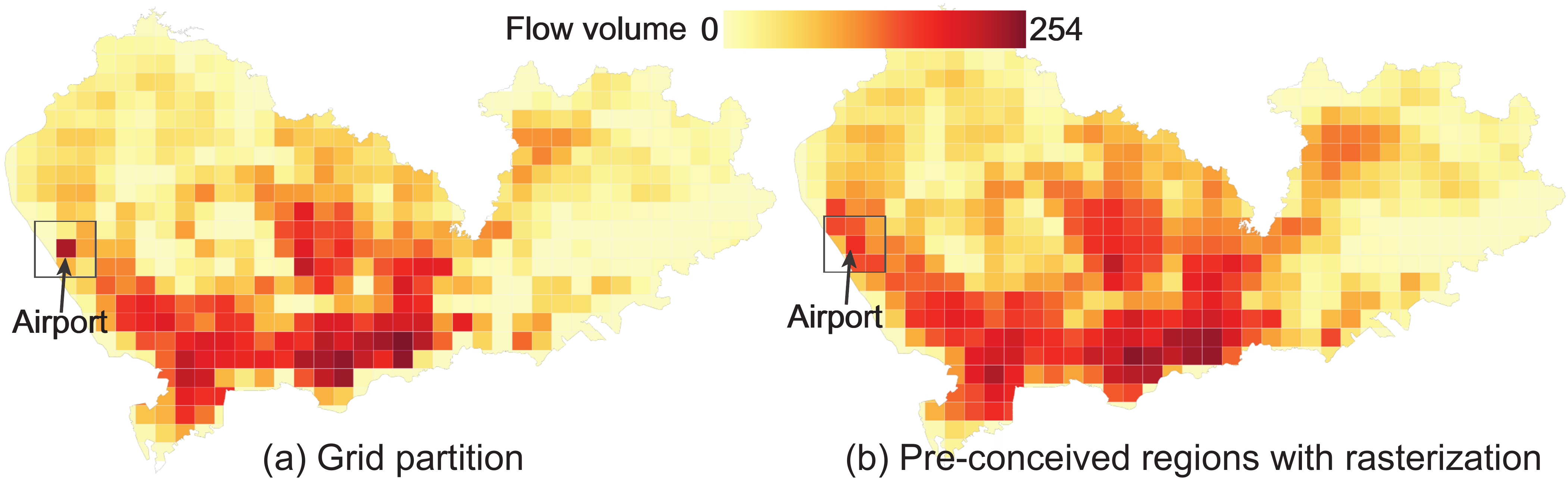}
  \vspace{-7mm}
  \caption{Comparison of traffic flows in Shenzhen by grid partition based on the longitude and latitude (a), and by pre-conceived regions with rasterization followed (b).
  Both raster images are with size of 50$\times$25.}
  \vspace{-4mm}
  \label{fig:grid_vs_raster}
\end{figure}

In this way, we construct a raster image $X_{G,t} \in \mathbb{R}^{i\times j}$ that represents the in/out traffic flows of the grid map $G$.
We perform the rasterization process for all time slots $t \in \mathcal{T}$, yielding a series of raster images $\left\{ X_{G, t} \in \mathbb{R}^{i \times j} | t \in \mathcal{T} \right\}$.
Fig.~\ref{fig:grid_vs_raster} presents a comparison of traffic flows in Shenzhen by grid partition of a 50$\times$25 grid map (a) and pre-conceived regions of 491 TAZs in Shenzhen (b). 
There are noticeable differences between the two raster images.
Especially in the airport area, grid partition produces an extreme high volume for the grid in the center, whilst the grid has similar flow volume with those surrounding grids in the pre-conceived regions.
The reason is that the TAZ where locates the airport is big, and the traffic flows are averaged by grids intersecting with the TAZ.
The input differences have a significant impact on prediction accuracy of the outputs, as we will show in the experiment.

\begin{figure*}
  \centering
  \includegraphics[width=0.995\textwidth]{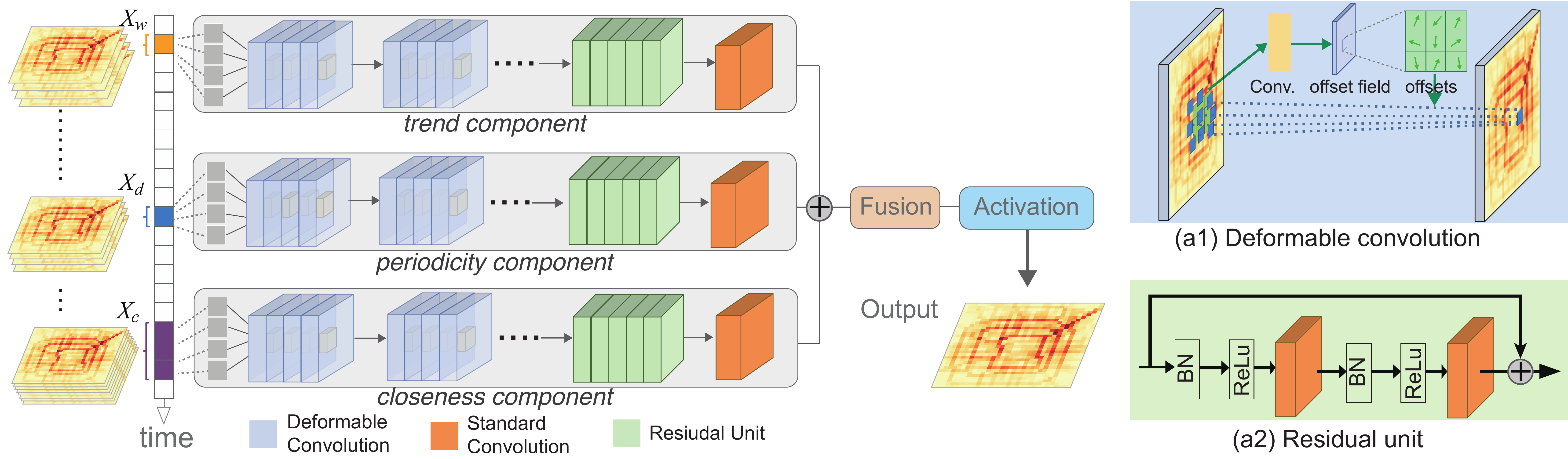}
  \vspace{-3mm}
  \caption{Network architecture of our \emph{DeFlow-Net}, which mainly consists of three modules: (i) a \emph{temporal dependency module} including weekly \emph{trend}, daily \emph{periodicity}, and hourly \emph{closeness} components to learn periodic patterns; (ii) a \emph{deformable convolution module} taking advantage of deformable convolutions (b1) to learn spatial dependence, and residual units (b2) to increase network depth; and (iii) a \emph{fusion and activation module} to fuse temporal components and activate the final prediction.}
  \vspace{-5mm}
  \label{fig:DeFlow-Net}
\end{figure*}

\subsection{DeFlow-Net}
\label{ssec:deflow-net}

\textbf{Overview}.
We design \emph{DeFlow-Net} that takes a series of raster images $\left\{ X_{G, t} \in \mathbb{R}^{i \times j} | t \in \mathcal{T} \right\}$ as inputs, and predicts traffic flows $Y_{G, t_{n+1}}$ for the grid map $G$ at time slot $t_{n+1}$.
The predicted traffic flow $Y_{G, t_{n+1}}$ can be expressed as: 

\begin{equation}
Y_{G, t_{n+1}} = \mathcal{F}_{\theta}(\left\{ X_{G, t_1}, X_{G, t_2}, \cdots, X_{G, t_n} \right\}),
\end{equation}
where $\mathcal{F}_{\theta}(\cdot)$ represents the proposed \emph{DeFlow-Net} model, while $\theta$ are the learnable parameters.
Fig.~\ref{fig:DeFlow-Net} presents the overall framework of \emph{DeFlow-Net}, which is composed of the following modules:
\begin{itemize}

\item
\emph{Temporal Dependency Module} (Sect.~\ref{sssec:temporal_module}).
The module models the periodic patterns of traffic flows using temporal components of weekly trend, daily periodicity, and hourly closeness.

\item
\emph{Deformable Convolution Module} (Sect.~\ref{sssec:convolution_module}).
The module learns spatial dependence and nonstationarity using deformable convolutions, and deepens the convolutional layers using residual unit.

\item
\emph{Fusion and Activation Module} (Sect.~\ref{sssec:fusion_module}).
Last, \emph{DeFlow-Net} fuses the three temporal components to model spatial-temporal correlation, and activates the final prediction.

\end{itemize}

\subsubsection{Temporal Dependency Module}
\label{sssec:temporal_module}

Traffic flows in historical data feature periodic patterns.
Most of the studies on short-term traffic flow predictions, including deep-learning-based approaches (\eg,~\cite{guo2019deep,zhang_2017_deep}), make use of this property. 
Following previous conventions, \emph{DeFlow-Net} employs three components of weekly \emph{trend} that associates distant time slots in one-week, daily \emph{periodicity} that associates near time slots in one-day, and hourly \emph{closeness} that associates recent time slots in a few hours, to model temporal dependency.

Inputs for weekly trend component are a subsequence of raster images from the previous week.
We can define the weekly trend component as $\left[X_{G, t-\Delta w \cdot p_{w}}, X_{G, t- (\Delta w-1) \cdot p_{w}}, \ldots, X_{G, t-p_{w}}\right]$, where $\Delta w$ represents length of the raster image subsequence from the previous weeks, and $p_{w}$ is fixed to one-week length, \ie, $p_{w} = 48 \times 7$.
Similarly, we define daily trend component as
$\left[X_{G, t-\Delta d \cdot p_{d}}, X_{G, t-(\Delta d-1) \cdot p_{d}}, \ldots, X_{G, t-p_{d}}\right]$, where $\Delta d$ represents length of the raster image subsequence from the previous days, and $p_{d}$ is fixed to one-day length, \ie, $p_{d} = 48$.
In the closeness component, we select the previous $\Delta c$ time slots of traffic flows to infer the next time slot.
Hence, the closeness component is defined as $\left[X_{G, t-\Delta c }, X_{G, t- (\Delta c-1)}, \ldots, X_{G, t-1}\right]$.

\subsubsection{Deformable Convolution Module}
\label{sssec:convolution_module}

Traffic flows that measure continuous movements across spatial units are by nature spatially dependent.
For instance, regions with similar functionality~\cite{yuan_2012_discovering} or POI distributions~\cite{zeng_2017_visualizing} are likely to show similar traffic flows as well.
We utilize a \emph{Deformable Convolution Module} to model spatial dependence among traffic flows.

Inputs of the module are \emph{trend}, \emph{periodicity}, and \emph{closeness} components from the \emph{Temporal Dependency Module}.
We first concatenate the sequential raster images in each component as one tensor.
Taking the \emph{closeness} component for example, we concatenate both inflow and outflow raster images $\left[X_{G, t-\Delta c }, X_{G, t- (\Delta c-1)}, \ldots, X_{G, t-1}\right]$ together, yielding $X_{G, c}^{(1)} \in \mathcal{R}^{2 \Delta c \times i \times j}$ (notation $^{(1)}$ indicates input for the first convolutional layer).
$X_{G, c}^{(1)}$ is then fed into multiple convolutional layers.
The transformation matrix for the \emph{l}-th convolutional layer can be defined as:

\begin{equation}
X_{G, c}^{(l+1)} = f_c(W_{G, c}^{(l)} \ast X_{G, c}^{(l)} + b_{G, c}^{(l)}),
\end{equation}
where $\ast$ denotes the convolution operation, while $W_{G, c}^{(l)}$ and $b_{G, c}^{(l)}$ are two sets of learnable parameters in the $k$-th convolution layer.
$f_c(\cdot)$ is the activation function, for which we use the rectified linear unit (ReLU), \ie, $f_c(z) = max(0, z)$.
To ensure the input and output have the same size, we use zero paddings for grids at the boundary.

\begin{figure}
  \centering
  \includegraphics[width=0.495\textwidth]{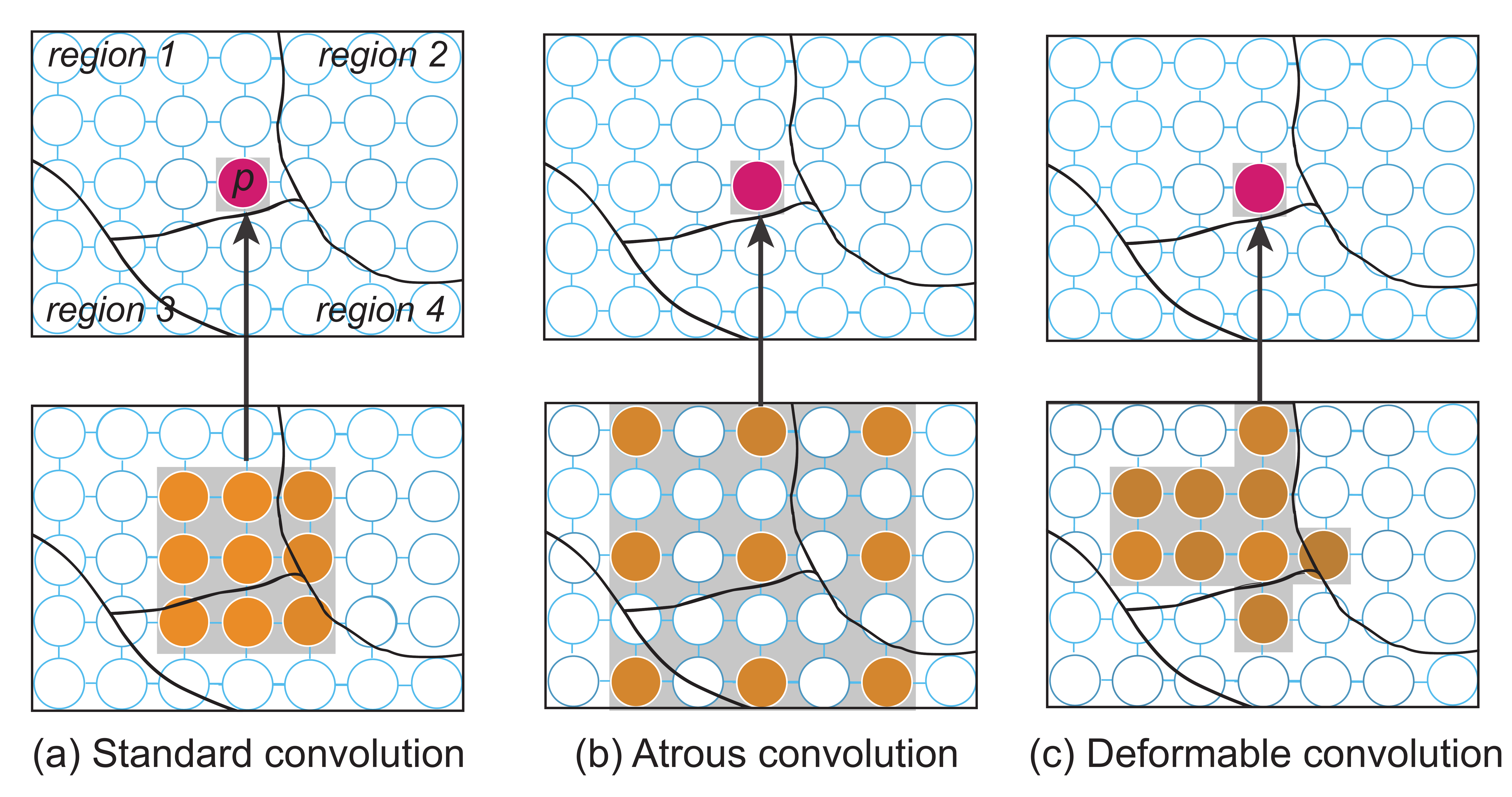}
  \vspace{-7mm}
  \caption{Illustration of the convolution alternatives. Left to right: (a) the receptive field of a standard convolution covers grids from different regions; (b) atrous convolution (atrous rate $r$ = 2) enlarges the receptive field but the grids are also from different regions; (c) deformable convolution covers neighboring grids or grids in the same region.}
  \vspace{-4mm}
  \label{fig:standard_vs_deformable}
\end{figure}

\vspace{1.5mm}
\noindent
\textbf{Convolutional Layers}.
Convolution is applied to make use of the spatial relationship among nearby grids covered by the receptive field.
Intuitively, convolutions can learn spatial dependence because traffic flows in neighboring regions can affect each other.
There are three convolution alternatives (see Fig.~\ref{fig:standard_vs_deformable}) that can lead to different effects:

\begin{itemize}

\item
\textbf{Standard convolution}.
Consider a location $\textbf{p}$ (red unit in the top row of Fig.~\ref{fig:standard_vs_deformable}) in $X_{G, c}^{(l+1)}$, a standard convolution can be defined as:

\begin{equation}
X_{G, c}^{(l+1)}\left[ \textbf{p} \right] = \sum_{k \in \mathcal{K}} W_{G,c}^{(l)} \left[ \textbf{k} \right] \ast X_{G,c}^{(l)} \left[ \textbf{p} + \textbf{k} \right],
\end{equation}

where $X_{G, c}^{l} \left[ \textbf{p} + \textbf{k} \right]$ represents pixel at the position $\textbf{p}+\textbf{k}$ in $X_{G, c}^{(l)}$, and $\textbf{k}$ is the location offset in $\mathcal{K}$.
For a kernel of the size $3\times3$, $\mathcal{K}$ is defined as: 

\begin{equation}
\mathcal{K}=\{(-1,-1),(-1,0), \ldots,(0,1),(1,1)\}.
\end{equation}

Standard convolution is commonly adopted in many CNN architectures.
However, standard convolution may not be optimal for learning spatial dependence in short-term traffic prediction.
As shown in Fig.~\ref{fig:standard_vs_deformable}(a), the unit marked in red is in region 1, whilst the receptive field of a standard convolution covers units distributed in 4 different regions.
If the regions have different functionalities, spatial dependence of traffic flows among these units would be low, causing weak feature representation.

\item
\textbf{Atrous convolution}.
Atrous convolution enlarges a convolution's field-of-view by inserting holes between pixels, to capture objects at multiple scales.
Atrous convolution can be defined as:

\begin{equation}
X_{G, c}^{(l+1)}\left[ \textbf{p} \right] = \sum_{k \in \mathcal{K}} W_{G,c}^{(l)} \left[ \textbf{k} \right] \ast X_{G,c}^{(l)} \left[ \textbf{p} + r \times \textbf{k} \right],
\end{equation}

where $r$ is the atrous rate that specifies $r-1$ zeros along each spatial dimension.
In case that $r = 1$, atrous convolution is equivalent to standard convolution.
As such, atrous convolution can be regarded as a general form of standard convolution.

Atrous convolution can improve the feature capturing capability with a large atrous convolution, as shown in Fig.~\ref{fig:standard_vs_deformable}(b).
However, the receptive field is still symmetric, forcing the convolutional layer to learn spatial dependence in symmetric units.
In contrast, regions are typically in irregular shapes.
In this sense, atrous convolution may even worsen the feature representation of spatial dependence, as it learns from distant regions~\cite{tobler1970computer}.

\item
\textbf{Deformable convolution}.
The deficiency of standard and atrous convolutions also applies to image segmentation of segmenting an image into multiple segments in regular shapes.
Deformable convolution is proposed to address the deficiency, by introducing additional 2D offsets to the receptive field~\cite{dai2017deformable}.
A deformable convolution can be expressed as:

\begin{equation}
X_{G, c}^{(l+1)}\left[ \textbf{p} \right] = \sum_{k \in \mathcal{K}} W_{G,c}^{(l)} \left[ \textbf{k} \right] \ast X_{G,c}^{(l)} \left[ \textbf{p} + \textbf{k} + \Delta \textbf{k} \right],
\end{equation}

where $\Delta \textbf{k}$ is the offset applied to the location offset $\textbf{k}$.
Fig.~\ref{fig:standard_vs_deformable}(c) shows an illustration of a deformable convolution applied to traffic flows. 
Here, the receptive field overlaps much with region 1, where $\textbf{p}$ is located.
The other two sampling units are outside region 1, but they are neighbors of $\textbf{p}$.
That is, the deformable convolution effectively learns spatial dependence from neighboring units, or those located in the same region with $\textbf{p}$.

As shown in Fig.~\ref{fig:DeFlow-Net}(b), the offset $\Delta \textbf{k}$ is obtained through a convolutional layer appended after a standard convolution layer.
In this way, \emph{DeFlow-Net} can realize the standard back-propagation for end-to-end training.
It is worth noting that learnable parameters $\Delta \textbf{k}$ are mostly fractional.
We complete the convolution by bilinear interpolation, in order to obtain the values at the fractional positions.
\end{itemize}

\vspace{1.5mm}
\noindent
\textbf{Residual Unit}.
Nevertheless, one convolutional layer can only account for dependence in nearby regions, due to the limited size of kernels.
To further enrich the spatial dependence between distant regions, \emph{DeFlow-Net} utilizes multiple deformable convolutional layers.
However, error back propagation becomes more difficult as the number of layers increases, causing the model degradation and vanishing gradient problems.
Residual network~\cite{he2016deep} is introduced to address the problem, by utilizing shortcut connections achieved with batch normalization, ReLU activation, and standard convolution, as illustrated in Fig.~\ref{fig:DeFlow-Net}(c).
The residual unit can be expressed as:

\begin{equation}
X_{G, c}^{(l+1)}=X_{G, c}^{(l)}+\mathcal{F}_r\left(X_{G, c}^{(l)}, W_{G, c}^{(l)}\right),
\end{equation}
where $\mathcal{F}_r(\cdot)$ denotes the residual function, and $W_{G, c}^{(l)}$ is the weight matrix set associated with the residual unit that needs to be learned.
Note here standard convolution is adopted instead of deformable convolution to reduce computational cost.
Prediction performance is not affected since spatial nonstationarity has been learned by deformable convolutional layers already.

Finally, we fuse the feature maps in each component through a standard convolution.

\subsubsection{Fusion and Activation Module}
\label{sssec:fusion_module}

After multiple convolutional layers and residual units in the \emph{Deformable Convolution Module}, we derive $X'_{G, c}$ for the \emph{closeness} component.
Similarly, we derive $X'_{G, d}$ and $X'_{G, w}$ for the \emph{periodicity} and \emph{trend} components, respectively.
Next, \emph{DeFlow-Net} employs a fusion layer to fuse the three components together, to simultaneously model the spatial and temporal correlations.
The fusion layer can be expressed as:

\begin{equation}
X_{G}^{Fusion}=W_{c} \circ X'_{G, c} +W_{d} \circ X'_{G, d}+W_{w} \circ X'_{G, w},
\end{equation}
where $W_{c}$, $W_{d}$, and $W_{w}$ are learnable parameters that adjust the degrees of \emph{closeness}, \emph{periodicity}, and \emph{trend} components, respectively.
We use the Hadamard product for $\circ$, which produces the sum of element-wise multiplication of two matrices.
Last, we employ an active function to predict traffic flow $ Y_{G, t_{n+1}}$ at time slot $t+1$ for the grid map $G$. The activation can be expressed as: 
\begin{equation}
 Y_{G, t_{n+1}}=tanh(X_{G}^{Fusion}),
\end{equation}
where $tanh$ is a hyperbolic tangent that ensures the output values are between -1 and 1.

\subsection{Training Scheme}
\label{ssec:training}

We normalize the training data to [-1, 1] using Max-Min normalization on the input datasets.
Then, we denormalize the predictions and compare them with the groundtruths.
We use the Adam optimizer to minimize mean squared error between the predicted traffic flow $ Y_{G, t_{n+1}}$ and the groundtruth $X_{G, t_{n+1}} $:
\newline
\begin{equation}
\mathcal{L}(\theta)=\| X_{G, t_{n+1}}-Y_{G, t_{n+1}}||_{2}^{2},
\end{equation}
where $\theta$ is learnable parameters in our model.

We implement \emph{DeFlow-Net} with Keras that uses TensorFlow as the backend.
\emph{DeFlow-Net} can be trained in an end-to-end manner via back-propagation.
All kernels of the convolutions are set to $3\times3$ in size.
The parameters for the three temporal components are set as: $\Delta w = 1$, $\Delta d = 1$, and $\Delta c = 3$, which empirically produces good results.
The batch size during the training is set to 32 and the learning rate is set to 0.001.
In addition, to obtain optimal model parameters and prevent overfitting, we perform the early-stopping strategy during training to control the number of epochs.
All experiments were conducted on a server (AMD Ryzen 7 2700 8-Core Processor $\times 16$, NVIDIA GeForce RTX 2080 GPU) with Linux operation system.
\section{Experiment}
\label{sec:Experiment}

\begin{table}
\centering
\caption{Statistics of the datasets used in the experiment.}
\vspace{-2.5mm}
\label{table:Dataset}
\begin{tabular}{|l|c|c|c|}
\hline
\textbf{Dataset}             & \textbf{TaxiBJ}          & \textbf{BikeNYC}       & \textbf{TaxiSZ}        \\ \hline \hline
Data type                    & Taxi GPS                 & Bike rent     & Taxi GPS      \\ \hline
Location                     & Beijing                  & New York      & Shenzhen      \\ \hline
Time period (days)           & 528                      & 183           & 181           \\ \hline
Time slot (minutes)          & 30                       & 60            & 30            \\ \hline
Available time slots           & 22,459                   & 4,392         & 8,688         \\ \hline \hline

Has TAZ?                     & No                       & No                            & 491 TAZs \\ \hline
\multirow{2}{*}{Grid map size} & \multirow{2}{*}{$32\times32$} & \multirow{2}{*}{$8\times16$}   &$50\times25$ \& \\
                             &                          &               & $100\times50$  \\ \hline
\end{tabular}
\end{table}

We conduct extensive experiments to evaluate the effectiveness of our proposed \emph{DeFlow-Net}.
This section presents the datasets (Sect.~\ref{ssec:DataSets}), baseline models and ablated techniques (Sect.~\ref{ssec:baselines}), and quantitative comparison results (Sect.~\ref{ssec:Result}).
In the end, we further perform spatial autocorrelation analysis (Sect.~\ref{ssec:spatial_analysis}), and discuss the impacts of different partition shapes and scales (Sect.~\ref{ssec:diffshape}).

\subsection{DataSets}
\label{ssec:DataSets}

We use three real-world datasets to evaluate the performance of our model: TaxiBJ, BikeNYC, and TaxiSZ.
Statistics of the datasets are summarized in Table~\ref{table:Dataset}, and the details are described as follows:

\begin{itemize}
\item \textbf{TaxiBJ:}
\emph{TaxiBJ} records taxicab GPS data in Beijing, during four periods from $1st$ $Jul$. 2013 to $30th$ $Oct$. 2013, $1st$ $Mar$. 2014 to $30th$ $Jun$. 2014, $1st$ $Mar$. 2015 to $30th$ $Jun$. 2015, and $1st$ $Nov$. 2015 to $10th$ $Apr$. 2016.
There are in total 528-day records divided into 30-minute time slots, yielding a total of 22,459 time slots (some slots are omitted due to corrupted data).
The city of Beijing is divided into a $32 \times 32$ grid map according to the longitude and latitude.
The data from the last four weeks are used for testing, while the rest are used for training.

\item \textbf{BikeNYC:}
\emph{BikeNYC} records bike trajectories taken from the the NYC Bike System, from $1st$ $Apr$. 2014 to $30th$ $Sept.$ 2014.
There are in total 183-days records divided into 60-minute tile slots, yielding a total of 4,392 time slots.
BikeNYC is organized in a grid map of scale $8 \times 16$.
We take the last ten days data for testing, and the rest for training.

\item \textbf{TaxiSZ:}
\emph{TaxiSZ} records taxi transactions carried out by over 20k taxis during the period from $1st$ $Jan$. 2019 to $30th$ $Jun$. 2019.
Unlike publicly available \textit{TaxiBJ} and \textit{BikeNYC} that have been cleaned up and preprocessed, \emph{TaxiSZ} contains approximately 800k raw transaction records per day, making a total of over 145 million transaction records in 181 days.
The raw data include numerous corrupted or incomplete information, such as positions outside Shenzhen or missing get-on/-off times.
We clean up the data by removing these records, yielding about 128 million transaction records ready for use.
Next, we divide the data into 30-minute time slots, resulting in a total of 8,688 time slots.
We use the data from the last two weeks for testing, and the rest for training.

Besides, we also have 491 TAZs that are designated by the transportation department in Shenzhen for tabulating traffic-related census data.
The TAZ borders correspond well with recognizable physical boundaries, such as main streets and waters. 
In this way, the land use and activities within each TAZ are relatively homogeneous.
We use the TAZ information to evaluate the benefits of partition by pre-conceived regions.

\end{itemize}

\subsection{Baselines and Ablations}
\label{ssec:baselines}
To evaluate the effectiveness of \emph{DeFlow-Net}, we first compare with five baseline models as follows:
\begin{itemize}
\item \textbf{HA}:
Historical Average (HA) uses historical average traffic flows of a given region at the corresponding time.
We average traffic flows of the last three week to make the estimation.

\item \textbf{ARIMA}:
Auto-Regressive Integrated Moving Average (ARIMA) is a well-known method for predicting future trends of a time series.
ARIMA has been widely used in traffic flow prediction.

\item \textbf{ST-ResNet}~\cite{zhang_2017_deep}:
ST-ResNet is one of the first convolution-based deep traffic flow prediction models.
ST-ResNet can also incorporate external factors such as weather information, which is omitted in the experiment for fair comparison.

\item \textbf{ST-3DNet}~\cite{guo2019deep}:
ST-3DNet exploits a specially designed 3D CNN architecture to learn spatial and temporal features in traffic flows simultaneously.
It is one of the latest convolution-based deep traffic flow prediction models.
\item \textbf{T-GCN}~\cite{zhao2019t}:
T-GCN learns to capture spatial dependencies from network topology with graph convolution, and temporal dependence from dynamic changes of traffic status with gated recurrent units.
It was originally proposed to predict traffic data on urban road network.
We model neighboring topology of the regions as a graph, and employ T-GCN to predict region-wise traffic flows.
\end{itemize}

We use the same training scheme for the baseline models, including loss function and kernel size, as those in \emph{DeFlow-Net}. 
Next, we also compare to two ablated techniques that utilize standard convolutions and atrous convolutions, instead of deformable convolutions in \emph{DeFlow-Net}.

\begin{itemize}

\item \textbf{Standard Convolution:}
We use a fixed kernel of $3 \times 3$ size to account for adjacency relationships.
All other settings are the same with those in \emph{DeFlow-Net}.

\item \textbf{Atrous Convolution:}
We use a $3 \times 3$ kernel with an atrous rate of 2.
This setting uses 9 parameters, but achieves the same field of view as a $5 \times 5$ standard convolution kernel.
All other settings are the same with those in \emph{DeFlow-Net}.
\end{itemize}

Last, we test the impacts of partition shapes and scales on prediction performances.
We compare grid, TAZ, and self-organized partition for \emph{TaxiSZ}, whilst omit \textit{TaxiBJ} and \textit{BikeNYC} that have been preprocessed using grid partition.

\begin{itemize}
  \item \textbf{TaxiSZ (Grid) 50$\times$25}: We first partition the underlying territory of Shenzhen into 50$\times$25 grids based on the longitude and latitude, then aggregate in- and out-flows in each grid.

  \item \textbf{TaxiSZ (Grid) 100$\times$50}: The city is partitioned into 100$\times$50 grids for aggregation of in- and out-flows.

  \item \textbf{TaxiSZ (TAZ) 50$\times$25}: We first aggregate traffic flows in each of the 491 TAZs, followed by a rasterization process using 50$\times$25 grid map.

  \item \textbf{TaxiSZ (TAZ) 100$\times$50}: The rasterization is performed upon a 100$\times$50 grid map after aggregation.

  \item \textbf{TaxiSZ (Self-Organized Region)}: 
  We generate two sets of regions $-$ one with 539 regions, and another one with 804 regions, by controlling the threshold as described in Equation~\ref{eq:similarity}.
  The basic units for constructing these self-organized regions are derived from a fine-grained demographic dataset with over 1k census blocks in Shenzhen. 
  For each set of regions, we rasterize into 50$\times$25 and 100$\times$50 grid maps.

\end{itemize}

We train a \emph{DeFlow-Net} model for each of these \emph{TaxiSZ} variants.
In the following, we abbreviate \emph{TaxiSZ (Grid) 50$\times$25} as \emph{TaxiSZ}, which is of similar size with TaxiBJ, when comparing to baseline models and ablated techniques.

\subsection{Performance Comparison}
\label{ssec:Result}

\textbf{Evaluation Metrics:}
To evaluate and compare the performance of different models, we first adopt root mean square error (RMSE) as in previous studies.
\begin{equation}
R M S E =\sqrt{\frac{1}{N} \sum_{g=1}^{N}\left({x}_{g,t_{n+1}}-y_{g,t_{n+1}}\right)^{2}},
\end{equation}
where $N$ is the number of grids, $x_{g,t_{n+1}}$ and $y_{g,t_{n+1}}$ are the ground truth and predicted traffic flow for grid $g$ at time slot $t_{n+1}$.
However, RMSE measurements are unit dependent, making it unsuitable for comparison between different datasets.
To address the problem, we further incorporate mean absolute scaled error (MASE)~\cite{hyndman2006another}.
\begin{equation}
M A S E=\frac{\frac{1}{N} \sum_{g=1}^{N}\left|x_{g, t_{n+1}}-y_{g, t_{n+1}}\right|}{\frac{1}{T-m} \sum_{g=1}^{N} \sum_{t=m+1}^{T}\left|x_{g, t}-x_{g, t-m}\right|},
\end{equation}
where $x_{g,t_{n+1}}$ and $y_{g,t_{n+1}}$ in the numerator are from the testing data, while $x_{g, t}$ and $x_{g, t-m}$ in the denominator are from the training data, respectively.
$T$ is the total number of time slots in the training data, and $m$ is the seasonality of the time series (\ie, 48 for \emph{TaxiBJ} and \emph{TaxiSZ}, and 24 for \emph{BikeNYC}).
MASE is unit independent, allowing us to compare traffic flow predictions in different cities and at different scales.
Moreover, MASE can handle actual values of zero and is not biased by very extreme values, which are problematic for mean absolute percentage error (MAPE)~\cite{hyndman2006another}.
In general, a MASE less than 1 indicates a model is better than the naive model, and lower MASE indicates better model.

\begin{table}[]
\begin{center}
\caption{Comparison with baseline models.}
\label{table:variants}
\setlength{\tabcolsep}{1.6mm}{
\begin{tabular}{c|c|c|c|c|c|c}
\hline
\multirow{2}{*}{Model} & \multicolumn{2}{c}{TaxiBJ} \vline      & \multicolumn{2}{c}{NYCBike} \vline  & \multicolumn{2}{c}{TaxiSZ} \\ \cline{2-7} 
                       & RMSE            & MASE            & RMSE             & MASE                  & RMSE          & MASE    \\ \hline
HA                     & 52.77           & 0.605           & 10.76            & 0.230               & 12.41         & 0.409   \\
ARIMA                  & 28.46           & 0.672                &  9.98            & 0.245              & 11.37         & 0.437       \\
ST-ResNet              & 17.34           & 0.295               & 6.48             & 0.171                 & 6.54          & 0.395       \\
ST-3DNet               & 17.14           & 0.292               & 5.95             & 0.168                 & 5.62          &  \textbf{0.234}      \\ 
T-GCN            & 39.68     & 0.633          &8.78            & 0.221 & 9.36 &  0.411       \\ \hline
DeFlow-Net             & \textbf{15.90}  &\textbf{0.278 }    & \textbf{5.85}    & \textbf{0.165}       & \textbf{5.35} & 0.236\\ \hline
\end{tabular}
}
\end{center}
\end{table}

\subsubsection{Comparison with Baselines}
\label{ssec:diffmodel}

Experimental results of comparison with baseline models are shown in Table~\ref{table:variants}.
The best performance for each dataset is marked in bold.
Here, conventional time-series models, \ie, \emph{HA} and \emph{ARIMA}, produce large RMSEs and MASEs.
This is probably because \emph{HA} and \emph{ARIMA} only make use of periodic patterns in temporal dimension, whilst deep-learning-based techniques further take advantages of spatial dependence.
Among the baselines (\emph{ST-ResNet}, \emph{ST-3DNet}, and \emph{T-GCN}), \emph{ST-3DNet} always achieves better performances for all datasets.
The result indicates that 3D convolutions can better learn spatial and temporal features, in comparison with 2D convolutions used in \emph{ST-ResNet}, and graph convolutions used in \emph{T-GCN}.
Specifically, \emph{T-GCN} achieves the worst performance among the deep learning models. 
A possible reason is that graph convolutions can only capture adjacent relationships among the regions, whilst neglects the spatial dependence in data with regular grid structure, \eg, traffic flows in regions.

On the other hand, our \emph{DeFlow-Net} is also based on 2D convolutions, but still achieves better performances in terms of RMSE and MASE for all experimental datasets. 
Specifically, our model reduces RMSEs to 15.90, 5.85, 5.35, and MASEs to 0.278, 0.165, 0.236, for TaxiBJ, NYCBike, and TaxiSZ respectively.
That is, \emph{DeFlow-Net} achieves on average 4.59\% RMSE and 1.91\% MASE improvements, in comparison with \emph{ST-3DNet}.
The result confirms the effectiveness of deformable convolutions that are employed in \emph{DeFlow-Net}, in modeling spatial characteristics of historical traffic flows.

\subsubsection{Comparison with Ablated Techniques}
\label{ssec:diffconv}

Table~\ref{table:diffconv} presents the comparison results of ablated techniques using standard and atrous convolutions.
We can notice that the ablation with atrous convolutions performs better in RMSE than the one with standard convolutions on all three datasets.
This is probably because atrous convolutions have a larger receptive field, allowing the ablation to better capture multi-scale spatial contextual information.
Yet the improvements in terms of MASE is marginal.
This indicates that the fixed geometric structure utlized by atrous convolutions, still limits its capability in modeling spatial nonstationarity of local traffic flows. 
Our model with deformable convolutions achieves better performances on all datasets.
Notice that the improvements on different datasets are slightly different.
In details, our model with deformable convolutions reduces RMSE by 7.99\%, 5.03\%, 7.60\%, and reduces MASE by 5.76\%, 3.51\%, 40.25\% than the ablation with atrous convolutions, for TaxiBJ, BikeNYC, and TaxiSZ, respectively.
The improvements on TaxiBJ and TaxiSZ datasets are more significant.
A possible reason is that the sizes of TaxiBJ and TaxiSZ grid maps are bigger, leaving more space for the model to learn sampling offsets.

\begin{table}[]
\caption{Comparison with ablated techniques of different convolutions.}
\label{table:diffconv}
\setlength{\tabcolsep}{1.6mm}{
\begin{tabular}{c|c|c|c|c|c|c}
\hline
\multirow{2}{*}{Convolution} & \multicolumn{2}{c}{TaxiBJ}  \vline & \multicolumn{2}{c}{BikeNYC} \vline & \multicolumn{2}{c}{TaxiSZ} \\ \cline{2-7} 
                             & RMSE         & MASE         & RMSE          & MASE         & RMSE         & MASE         \\ \hline 
Standard                     & 17.34        & 0.295        & 6.48          & 0.171        & 6.54         & 0.395        \\
Atrous                       & 17.22        & 0.295        & 6.16          & 0.175       & 5.79         &0.477           \\
Deformable                   & \textbf{15.90}& \textbf{0.278} &\textbf{5.85}   & \textbf{0.165} & \textbf{5.35}  & \textbf{0.236}    \\ \hline
\end{tabular}
}
\end{table}

\begin{figure*}
  \centering
  \includegraphics[width=0.995\textwidth]{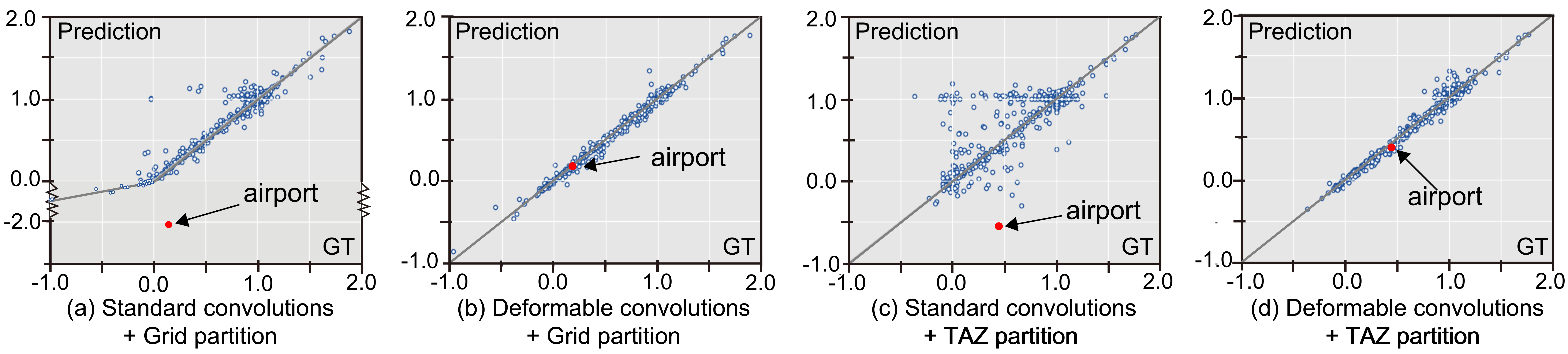}
  \vspace{-2mm}
  \caption{Q-Q plots of LISA indicators of groundtruth (x-axis) and predicted (y-axis) traffic flows by different combinations of spatial partition shapes and convolution types. Deformable convolutions can better preserve spatial autocorrelation, as the points in (b) \& (d) fall close to the diagonal line, whilst many points in (a) \& (c) are far distributed from the diagonal line.}
  \label{fig:local_moran}
\end{figure*}

\subsection{Spatial Autocorrelation Analysis}
\label{ssec:spatial_analysis}

Spatial autocorrelation is an important term in spatial statistics, which describes variation of a variable within geographic space.
Positive spatial autocorrelation indicates the tendency for locations that are close together to have similar values.
There exist many indicators for spatial autocorrelation analysis, such as \emph{Moran's I}~\cite{moran_I} and \emph{Geary's C}~\cite{geary_1954_contiguity}.
Both metrics are one single numeric statistic for measuring global spatial autocorrelation.
To quantify local effects that manifest spatial nonstationarity, this work adopts local indicators of spatial association (LISA) analyses that decompose the global Moran's I statistic to local Moran's I indices.

\begin{definition}[\textbf{Local Indicators of Spatial Association}]
Consider an in-/out-flow matrix $X_{G, t}$, we conduct LISA analysis for a local unit $x_{g_i, t}$ as: 
\begin{equation}
I\left(x_{g_{i}, t}\right)=\frac{x_{g_{i}, t}-\bar{X}_{G, t}}{S^{2}} \sum_{j=1, j \neq i}^{n} w_{i j}\left(x_{g_{j}, t}-\bar{X}_{G, t}\right)
\end{equation}
where $x_{g_{i}, t}$ indicates flow volume of grid $g_i$, and $x_{g_{j}, t}$ is the flow volume of grid $g_{j}$ that is one neighbor of grid $g_{i}$ at time slot $t$.
Here, we choose $3\times3$ surrounding grids as neighbours, corresponding to the size of receptive fields.
$w_{ij}$ is an element of spatial weights matrix that measures spatial connectivity between grids $g_i$ and $g_j$.
$\bar{X}_{G, t}$ is the mean flow volume of all grids in $G$, while $S^{2}$ is the variance of flow volumes of neighboring grids.
\end{definition}

A positive value of $I\left(x_{g_{i}, t}\right)$ indicates that grid $g_i$ has similarly high or low traffic volume as its neighbours, whilst a negative value indicates that $g_i$ is a spatial outlier.
We also conduct LISA analyses for the predicted traffic flows in the same way.
Fig.~\ref{fig:local_moran} presents Q-Q plots for LISA indicators of groundtruth (x-axis) and predicted (y-axis) traffic flows by (a) standard convolutions + grid partition, (b) deformable convolutions + grid partition, (c) standard convolutions + TAZ partition, and (d) deformable convolutions + TAZ partition.
All grid maps are of the size $50\times25$.
We can observe that grid partition produces many spatial outliers which have less than zero LISA indicators of the groundtruth traffic flows in (a) \& (b).

More importantly, we would like to examine and compare standard and deformable convolutions in terms of preserving spatial autocorrelation.
To do so, we further draw a diagonal line that passes through the coordinate origin and forms the 45 degree angle with the positive direction of the x-axis in each subfigure, as the reference.
Ideally, all points shall fall along the diagonal line, indicating LISA indicators of groundtruth and predicted traffic flows are the same for all grids.
By comparing standard and deformable convolutions, we can notice that deformable convolutions can better preserve spatial autocorrelation, as the points in (b) \& (d) fall close to the diagonal line, whilst many points in (a) \& (c) are far distributed from the diagonal line.
Taking the airport as an example, the grid is far away from the diagonal line in in (a) \& (c), whilst it is just besides the diagonal line in (b) \& (d).
This indicates that the ablated techniques using standard convolutions mess up predictions for the airport and neighboring grids, which may mislead users (\eg, taxi drivers) to wrong places when picking up passengers.

\begin{table}[]
\centering
\caption{Performance comparison of different partition shapes (grid \emph{vs.} TAZ) and scales ($50\times25$ \emph{vs.} $100\times50$) on TaxiSZ.}
\label{table:diffshape}

\begin{tabular}{c|c|c|c|c|c} \hline
\multirow{2}{*}{Convolution} & \multirow{2}{*}{Metric} & \multicolumn{2}{c|}{$50\times25$} & \multicolumn{2}{c}{$100\times50$}            \\ \cline{3-6} 
                             &                         & Grid & TAZ & Grid & TAZ \\ \hline
\multirow{2}{*}{Standard}    & RMSE                    & 6.54 & 6.81& 3.09 & 2.41   \\
                             & MASE                    & 0.395& 0.424 & 0.354 & 0.412 \\ \hline
\multirow{2}{*}{Atrous}      & RMSE                    & 5.79 & 5.73         & 2.90  & 2.27  \\
                             & MASE                    & 0.477& 0.461       & 0.344 & 0.422 \\ \hline
\multirow{2}{*}{Deformable}  & RMSE                    & 5.35 & 3.09 & 2.63 & 2.12 \\
                             & MASE                    & 0.236 & 0.322 & 0.329 & 0.347 \\ \hline
\end{tabular}
\end{table}

\subsection{Impacts of Partition Shape and Scale}
\label{ssec:diffshape}

We further evaluate the performance of \emph{DeFlow-Net} upon various partition schemes.
Here we divide the underlying territory into different shapes (grids \emph{vs.} TAZs \emph{vs.} self-organized regions) and scales ($50\times25$ \emph{vs.} $100\times50$).
We also compare the results with ablated techniques using standard and atrous convolutions.
Table~\ref{table:diffshape} presents the experiment results for grid and TAZ partition, which reveal some interesting findings:

\begin{itemize}
\item
First, our proposed \emph{DeFlow-Net} outperforms the ablated techniques in all partition schemes, in terms of both RMSE and MASE.
Notice that at partition scale $100\times50$, the improvements by \emph{DeFlow-Net} is relatively small than that at scale $50\times25$.
One possible reason is that at the finner scale, traffic flows are small for most of the grids, making spatial stationary across geographic space.
In such case the benefits of deformable convolutions drop.

\item
Second, finer scale (\ie, $100\times50$) always produces better results than coarse scale (\ie, $50\times25$) in terms of RMSE.
The results however do not infer that finer scale is better, but simply because RMSE is a unit-dependent metric. 
In contrast, the unit-independent metric MASE recommends coarse scale $50\times25$.
Future researches shall take care about the unit dependence issue.

\item
Last, at the same partition scale, \emph{DeFlow-Net} with TAZ partition always produces better results than with grid partition in terms of RMSE, but not of MASE.
Through careful investigation, we find the higher MASEs by TAZ partition mainly come from some grids at the boundary regions.
The rasterization process assigns some marginal traffic flows to these grids, and eventually cause large MASEs.

\end{itemize}

\begin{table}[]
\centering
\caption{Performance comparison of self-organized partition with varying number of regions (539 \emph{vs.} 804) and scales ($50\times25$ \emph{vs.} $100\times50$).}
\label{table:self-organize}
\begin{tabular}{c|c|c|c|c}
\hline
Num. Regions  & \multicolumn{2}{c|}{539 ($\alpha$ = 0.01)}  & \multicolumn{2}{c}{804 ($\alpha$ = 0.02)} \\ \hline
Grid map    & 50$\times$25 & 100$\times$50 & 50$\times$25 & 100$\times$50 \\ \hline
RMSE & 3.07  & 2.03   & 3.68  & 2.23   \\
MASE &0.323  & 0.330  & 0.415 & 0.443  \\ \hline
\end{tabular}
\end{table}

Table~\ref{table:self-organize} further presents the performance comparison of self-organized partition with 539 and 804 regions made by different threshold values.
We also compare the results under different grid map size ($50\times25$ \emph{vs.} $100\times50$).
All the results are achieved by \emph{DeFlow-Net} with deformable convolutions.
Here, the prediction accuracy measured by both RMSE and MASE improves when the final number of regions decreases from 804 to 539.
A possible reason is that traffic flows are averaged when merging adjacent regions, \ie, spatial dependence increases as the number of regions decreases.
Nevertheless, the results do not infer that less regions are better, as it would be practically more useful to predict traffic flows for fine-grained regions.
Hence, we stop merging regions when the number reaches 539, which is similar to the number of regions (491) of TAZ partition.
Moreover, we can notice that self-organized partition with 539 regions achieves a slightly better performance than TAZ partition with 491 regions (Table~\ref{table:diffshape}), even though the number of regions by self-organized partition is more.
This is probably because that self-organized partition based on traffic flows can better accord with spatial autocorrelation as compared to TAZ partition.
\section{Related Work}
\label{sec:related_work}

This work focuses on short-term forecasting for traffic flows~\cite{vlahoginani_2004_short-term}.
We group related work into three categories:
\emph{Traffic flow prediction} summarizes conventional and lately deep-learning-based methods for traffic flow prediction;
\emph{spatial nonstationarity} discusses spatial nonstationarity of traffic flows, and its impacts on deep traffic flow prediction;
and \emph{deformable convolutions} introduces recent developments of deformable convolution networks.

\subsection{Traffic Flow Prediction}

Traffic flow prediction can be regarded as a classic time-series forecasting problem.
As such, many conventional methods for traffic flow prediction are based on time-series models, such as autoregressive integrated moving average (ARIMA) (\eg,~\cite{moorthy_1988_short, willianms_1998_urban}) and structural time-series model (STM)~\cite{ghosh_2009_multivariate}, which take advantage of repeating occurrences in temporal historical data to fit parametric models.
An alternative approach is nonparametric regression, including \emph{k}-nearest neighbors (\emph{k}-NN) (\eg,~\cite{chang2012dynamic, wu2004travel}) and Bayesian network (\eg,~\cite{sun_2006_bayesian}).
Comparison study~\cite{smith_2002_comparison} showed that parametric models generally outperform nonparametric regressions, yet the performance of nonparametric regressions can be significantly improved by larger databases.
For instance, a large value of $k$ coupled with larger databases can provide a better set of neighbors in $k$-NN models, and consequently improve the prediction accuracy.

Nevertheless, the complexity of finding neighbors increases dramatically with larger $k$ values and database sizes.
Deep neural networks (DNNs) can suppress the challenge by learning hidden features in traffic flows via dedicated network architectures. 
As a result, DNNs are being increasingly used for traffic flow prediction.
Early attempts with fully connected neural networks~\cite{huang2014deep, lv_2015_traffic} achieve superior performances than conventional approaches.
Lately, CNNs (\eg,~\cite{zhang_2017_deep, du2019deep, zhang2019flow, gong_2020_sd-seq2seq}) and GNNs (\eg,~\cite{yan_2018_st-gcn, zheng2020gman, guo2019attention,zhao2019t}), have been utilized for predicting traffic flows.
CNNs model spatial dependency by decomposing the traffic network as grids, while GNNs are more appropriate for graph-structured traffic data, \eg, the road network~\cite{ye_2020_how}.

This work focuses on predicting traffic flows in regions, for which CNNs are more frequently utilized.
The experimental results (Sect.~\ref{ssec:diffmodel}) have shown that both 2D and 3D CNNs outperform \emph{T-GCN} using graph convolution networks.
This is probably because graph convolutions can only capture adjacent relationships among the regions, but not the spatial dependence and nonstationarity properties among regions.

However, existing convolution-based methods typically adopt standard convolutions with fixed geometric structures to learn spatial dependence across the whole space.
The assumption of spatial stationarity is problematic, since a global trend do not reflect the underlying data generating processes~\cite{brunsdon1996geographically}.
Instead, we employ deformable convolutions that are competent to model the sophisticated spatial nonstationarity in local traffic flows.

\subsection{Spatial Nonstationarity}
When modeling spatial-temporal dynamics, Anselin emphasized in his influential textbook~\cite{anselin2013spatial} two intrinsic characteristics of spatial data that need to be taken into account: \emph{spatial dependence} and \emph{spatial nonstationarity} (or spatial heterogeneity).
As Tobler's first law of geography~\cite{tobler1970computer} observed, ``everything is related to everything else, but near things are more related than distant things", \emph{spatial dependence} refers to a situation where attribute values observed at one spatial unit are dependent on neighboring values at nearby spatial units.
In contrast, \emph{spatial nonstationarity} points to the lack of spatial stationarity in attribute values of a particular measure across all spatial units~\cite{brunsdon1996geographically}.

By far, the literature of convolution-based deep traffic flow prediction concentrates on modeling spatial dependence, whilst little attention has been put on modeling spatial nonstationarity.
Yao et al.~\cite{yao2018deep} showed that including regions with weak correlations for a target region can hurt the prediction performance, and designed local CNNs to filter weakly correlated remote regions.
Cheng et al.~\cite{Cheng_2020_short} proposed a dynamic spatio-temporal \emph{k}-nearest neighbor model to capture the heterogeneous spatio-temporal pattern of road traffic.
Their work is inspiring but differs with our approach that utilizes deformable convolutions to \emph{automatically} learn spatial nonstationarity from traffic flows.
Perhaps most similar to our work is DST-ICRL~\cite{du2019deep} that combines irregular convolutional residual network with long short term memory network.
However, DST-ICRL needs to divide passenger flows of different traffic lines or routes into multiple channels, and utilizes irregular convolution kernels to capture the interpretable high-level dependency in each individual channel.
The process is time-consuming and may damage the dependence among spatial units. 
Instead, our proposed \emph{DeFlow-Net} reserves traffic flows of all spatial units as one single input, being able to learn spatial dependence and nonstationarity simultaneously.

\subsection{Deformable Convolution}

Convolutional layers that learn abstract feature maps using convolution kernels, are basic components in convolutional neural networks.
Standard convolution kernels are defined by fixed shapes of equal width and height (\eg, 3$\times$3 or 5$\times$5), which however may lose some neighboring information as the receptive field of a standard convolution kernel only covers an area with checkerboard patterns~\cite{wang_2018_understanding}.
The deficiency may impede certain prediction tasks that feature dynamic spatial correlations, such as semantic segmentation and traffic flow prediction.

Atrous convolution, also known as dilated convolution, inserts holes (\ie, \emph{trous} in French) between pixels to enlarge the field of convolution kernels~\cite{chen_2017_rethinking}.
A well-known example is DeepLab~\cite{chen_2018_deeplab} that employs Atrous Spatial Pyramid Pooling (ASPP) to capture multi-scale objects and context information by placing multiple dilated convolution layers in parallel.
Although atrous convolution can enable dense feature extraction, the convolutional kernels are designed to sample the input tensors at symmetric positions, making it difficult to align key points or salient features at arbitrary positions.
Therefore, atrous convolutions can still lead to the loss of total information given the discontinuous sampling~\cite{xu2020deformable}.

To overcome the limitation, Dai et al.~\cite{dai2017deformable} proposed deformable convolution as a general form of atrous convolution, by learning sampling offsets dynamically from input tensors.
Due to its generalizability, deformable convolution has been successfully applied in many computer vision applications, \eg, semantic segmentation~\cite{CHEN2021853}, image deblurring~\cite{yuan_2020_efficient}, and video enhancement~\cite{deng2020spatio}.
We introduce deformable convolutions for traffic flow prediction, and design \emph{DeFlow-Net} accordingly. Experimental results demonstrate the superior performance of deformable convolutions than standard and atrous convolutions.

\section{Discussion, Conclusion, and Future Work}
\label{sec:conclusion}

Though being an intrinsic property, the local nonstationary characteristic of traffic flows has been seldom considered in deep traffic flow prediction.
This work calls for attention to the impacts of spatial nonstationarity on convolution-based DNN models.
We tackle the problem from the perspectives of network inputs and architecture. 

First, we show that spatial partition schemes, including both partition shapes and scales (\ie, the MAUP~\cite{gehike_1934_certain, openshaw_1984_modifiable}), can affect local nonstationarity across the geographic space.
Consequently, prediction performance of CNN models are different.
Specifically, spatial partition according to pre-conceived regions (TAZs) and self-organized regions tend to relax spatial nonstationarity and increase prediction accuracy, in comparison to grid partition based on the longitude and latitude.
Besides, finer-grained partition scale (TaxiSZ $100\times50$) always produces smaller absolute errors than coarse scale (TaxiSZ $50\times25$), yet the inference is not reliable if we consider unit independent metrics, \eg, MASE adopted in this work.
Hence, we suggest

\noindent
\textbf{Insight 1:}
\emph{to integrate pre-conceived or self-organized regions} with traffic flows when preparing network inputs, and choose appropriate partition scale according to practical requirements.

\vspace{1.5mm}
Second, realizing the impacts of spatial nonstationarity, we design \emph{DeFlow-Net} (Sect.~\ref{ssec:deflow-net}), a deep deformable convolutional residual network that incorporates deformable convolutions to enhance the capability of extracting spatial features.
\emph{DeFlow-Net} outperforms existing convolution-based DNN models and ablated techniques, throughout all three real-world datasets.
The advancements include both prediction accuracy in terms of RMSE and MASE, and the ability to preserve spatial autocorrelation.
The results indicate that deformable convolutions can model both globally spatial dependence and locally spatial nonstationarity.
As such, we recommend

\noindent
\textbf{Insight 2:}
\emph{
to use deformable convolutions, instead of standard convolutions, when constructing the network architecture of convolution-based DNN models for deep traffic flow prediction.}

\vspace{1.5mm}
\noindent
\emph{Future work}:
There are several directions for future research, as follows.
First, the DNN models used in this work are treated as as `black box', and we apply `what-if' analysis to examine the impacts of spatial nonstationarity on prediction performance.
To better understand the working mechanism and improve the model, we would like to open the black box and display the internal states of the DNN models, such as to interpret the hidden unit response to feature selections~\cite{shen_2020_visual}.
Second, this work relies solely on traffic flows for prediction.
Studies have shown that incorporating external factors, such as weather~\cite{wang_2017_deepsd, zhang_2017_deep} and context information~\cite{Lin_Feng_Lu_Li_Jin_2019}, can improve the prediction performance.
We would also like to redesign the network architecture to fuse these factors.
Nevertheless, the heterogeneous data also exhibit independent spatial nonstationarity, bringing in more challenges for network prediction.

\section*{Acknowledgment}
This work is supported partially by National Natural Science Foundation of China (61802388, 62077039, 41901391), Guangdong Basic and Applied Basic Research Foundation (2021A1515011700), and the Shenzhen Fundamental Research Program (JCYJ20190807163001783).

\bibliographystyle{abbrv}
\bibliography{Reference}

\begin{IEEEbiography}[{\includegraphics[width=1in,height=1.25in,clip,keepaspectratio]{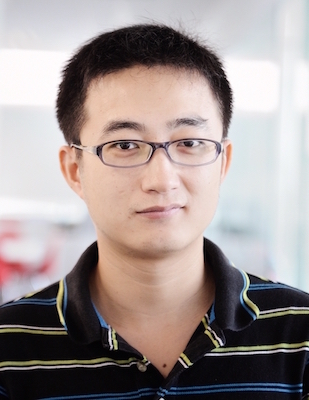}}]
{Wei Zeng} is currently an associate professor at Shenzhen Institute of Advanced Technology, Chinese Academy of Sciences.
He received both his B.E. and Ph.D. in computer science from Nanyang Technological University.
Before joining SIAT, he worked as a senior researcher at Future Cities Laboratory, ETH Zurich.
His current research interests include geospatial data analysis and visualization, AR/VR, and urban computing.
\end{IEEEbiography}

\begin{IEEEbiography}[{\includegraphics[width=1in,height=1.25in,clip,keepaspectratio]{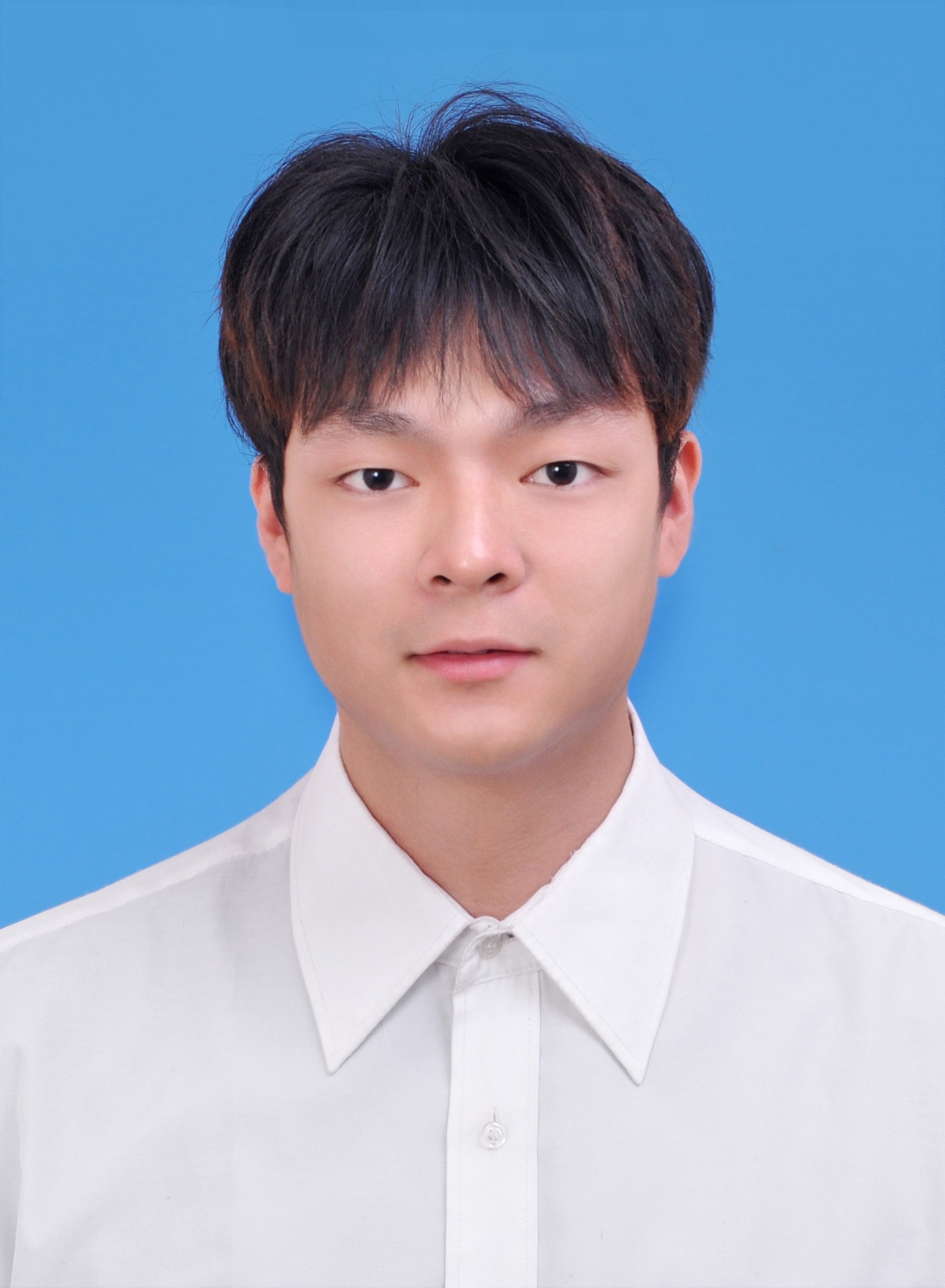}}]
{Chengqiao Lin} is currently a master student in School of Informatics, Xiamen University. His research interests include spatiotemporal data mining, intelligent transportation systems and urban computing.
\end{IEEEbiography}

\begin{IEEEbiography}[{\includegraphics[width=1in,height=1.25in,clip,keepaspectratio]{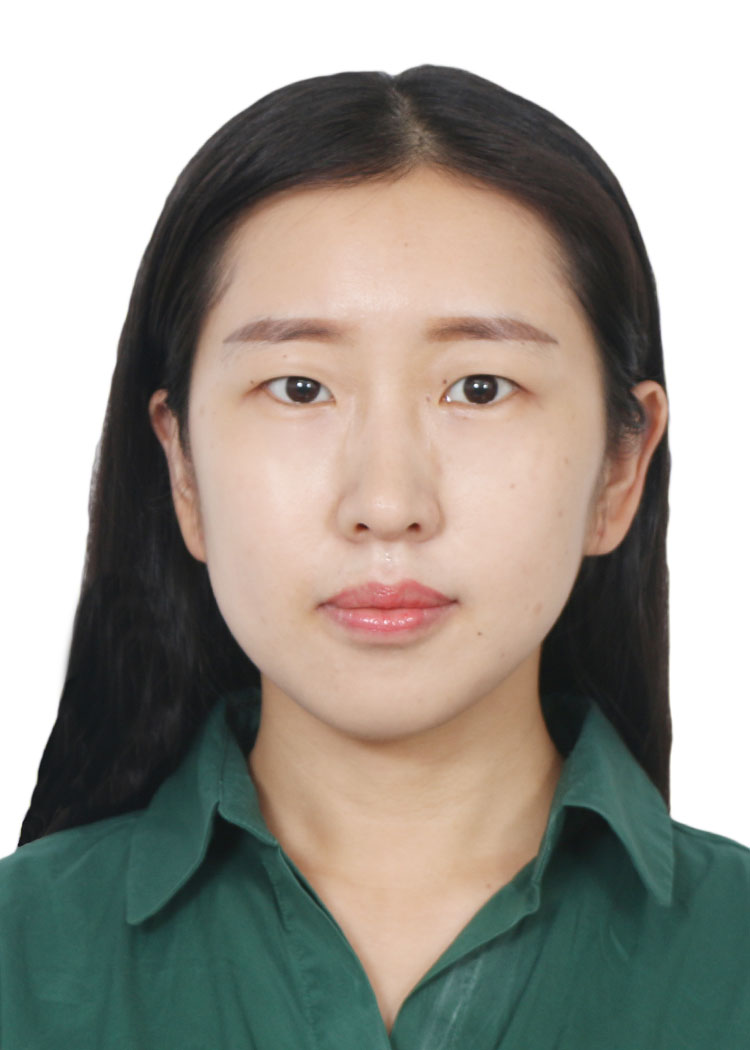}}]
{Kang Liu} is currently an associate professor at Shenzhen Institute of Advanced Technology, Chinese Academy of Sciences.
She obtained her Ph.D. degree in Institute of Geographic Sciences and Natural Resources Research (IGSNRR), Chinese Academy of Sciences (CAS) in 2018.
Her research interests focus on geospatial data intelligence and urban computing, and application fields of intelligent transportation, public health, and urban planning. 
\end{IEEEbiography}

\begin{IEEEbiography}[{\includegraphics[width=1in,height=1.25in,clip,keepaspectratio]{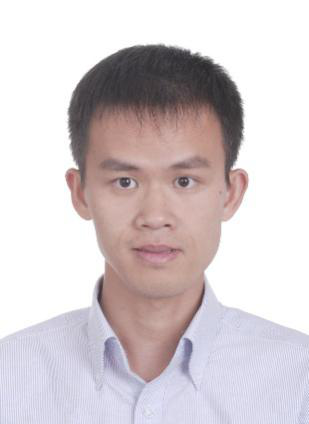}}]
{Juncong Lin} is currently a full professor in School of Informatics, Xiamen University, leading the Graphics and Virtual Reality Laboratory. He received his B.S. and Ph.D Degree both from Zhejiang Univesity in 2003 and 2008 respectively. Before joining Xiamen University, he has worked in CUHK, NTU as a postdoc researcher. His research interests include data analysis, educational visualiazation, creativity support tools, and geometry process.
\end{IEEEbiography}

\begin{IEEEbiography}[{\includegraphics[width=1in,height=1.25in,clip,keepaspectratio]{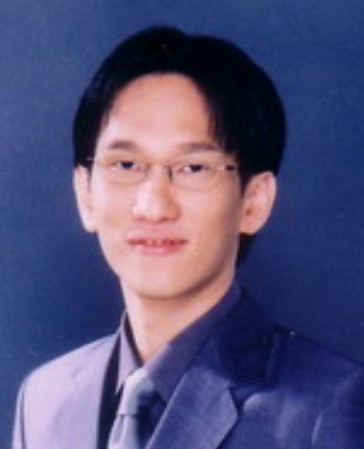}}]
{Anthony K. H. Tung} is currently a Professor in the Department of Computer Science, National University of Singapore (NUS).  He received both his B.Sc. and M.Sc. in computer sciences from the National University of Singapore in 1997 and 1998 respectively.  In 2001, he received the Ph.D. in computer sciences from Simon Fraser University (SFU).  Dr Anthony Tung main research areas are on searching, mining and visualizing complex data.  More recently, he also looks into the creation of innovative big data applications over the data processing techniques that he had developed over the past 18 years.  Anthony is also the deputy director of NUS NCript research center (\url{https://ncript.comp.nus.edu.sg/}).
\end{IEEEbiography}
\vfill


\end{document}